\documentclass[showpacs,preprintnumbers,amsmath,amssymb]{revtex4}

\usepackage{graphicx}
\usepackage{dcolumn}
\usepackage{bm}

\begin{document}


\title{Comparison of localization procedures for applications in crystal
embedding}

\author{O. Danyliv}
 \email{oleh.danyliv@kcl.ac.uk}
 \altaffiliation[On leave from]  { Institute for
Condensed Matter Physics, National Academy of Science of Ukraine,
Ukraine}
\affiliation{Department of Physics, Kings College
London, Strand, London WC2R 2LS, UK}%
\author{L. Kantorovich}%
 \email{lev.kantorovich@kcl.ac.uk}
\affiliation{Department of Physics, Kings College
London, Strand, London WC2R 2LS, UK}%

\date{\today}

\begin{abstract}
With the aim of future applications in quantum mechanical
embedding in extended systems such as crystals, we suggest a
simple and computationally efficient method which enables
construction of a set of nonorthogonal highly localized
one-electron orbitals for periodic nonmetallic crystals which
reflect their chemical nature. The orbitals are also used to build
up the Hartree-Fock (HF) electron density of the entire crystals.
The simplicity of the method stems from the fact that it does not
require usage and/or modification of periodic electronic structure
codes, and is instead based on the HF calculation of a sequence of
finite clusters with subsequent application of a localization
procedure to transform the HF canonical molecular orbitals. Two
extreme cases of chemical bonding, ionic (MgO crystal) and
covalent (Si crystal), are considered for which a number of known
localization schemes are applied and compared. With some
modifications our method can also be applied to nonperiodic
nonmetallic systems as well.
\end{abstract}

\maketitle

\section{Introduction}

Electronic structure calculations of extended systems with a local
perturbation, such as point defects in the bulk of crystals
\cite{Marshall-book} or adsorption of molecules at their surfaces
\cite{Chemisorption-Reactivity-1997} are of fundamental importance
in solid state physics and chemistry. Over the last decade a
number of effective computational techniques have been developed
to study the electronic ground state of such systems which are
based on periodic boundary conditions (PBC) and either
Hartree-Fock (HF) \cite{CRYSTAL98,CRYSTAL} or Density Functional
Theory (DFT) \cite{Payne-DFT,Parr-Yang} approaches. In these
methods a local perturbation (e.g. an adsorbed molecule together
with a fragment of a crystal surface) is artificially periodically
repeated in a cell which is large enough to ensure that the
interaction between periodic images is negligible. These methods
can also be applied to study extended (but not infinite)
biological systems in which important chemistry is usually
associated with a local part of the entire molecule(s)
\cite{Segall-2002}.

Another set of methods, commonly referred to as embedding
techniques, originate from a model in which a single local
perturbation is considered in the direct space of the entire
system. This makes the model closer to reality at low
concentration, but, at the same time, it makes it more challenging
since, due to the lack of periodic symmetry, well developed PBC
based techniques cannot be applied here. Instead, a number of
\emph{hybrid} methods have been developed which treat different
parts of the system at different levels of the theory. Most of
these methods combine \emph{ab initio} quantum mechanics (QM)
methods (based either on the DFT or the HF methods and their
extensions) applied to a finite fragment of the system (a
\emph{quantum cluster}), with molecular mechanics (MM) methods
based on semi-classical force fields and applied to the rest of
the system (\emph{environment region}). The idea is to consider
the most relevant part of the system (e.g. with respect to a
process in question) in great detail, while the rest of the system
is treated at a substantially lower level. These methods,
developed mostly within the quantum chemistry community, and
usually referred to as QM/MM have proven to be extremely
successful
\cite{QM/MM,Sauer-Sierka-2000,Hall-Hinde-Burton-Hillier-2000,Rivail3,Murphy-Philipp-Freisner-2000}.
Note that some other embedding schemes
\cite{EMC-1,EMC-2,Barandiaran-1996,Bredow-1999,Petja-2000,Sulimov2002,Nasluzov2001}
developed mostly in the solid state community are very close in
spirit to the QM/MM methods.

Almost in all embedding methods mentioned above the quantum
cluster is surrounded by point charges of the MM region. In
addition, in covalent systems the bonds coming out of the cluster
are usually terminated by pseudoatoms (see, e.g.
\cite{Sulimov2002}), so-called link atoms
\cite{Rivail3,Sauer-Sierka-2000,Murphy-Philipp-Freisner-2000} or
pseudopotentials
\cite{Abarenkov-Tupitsyn-2001,Abarenkov-Tupitsyn-2001r,Petja-2000,Nasluzov2001,Sulimov2002}.

Another class of embedding schemes relies on a more
{}``electronic'' (less {}``mechanical'') representation of the
environment region surrounding the quantum cluster. For instance,
in
\cite{Vreven-Morokuma-2000,Sauer-Sierka-2000,Abarenkov-Bulatov-1997}
this is achieved by a special total energy construction which
allows a combination of several electronic structure methods of
different complexity applied to different parts of the system;
other methods \cite{EMC-1,EMC-2,Barandiaran-1996} rely on a
representation of the wavefunction of the whole system as an
antisymmetrised product of strongly orthogonal many-electron group
functions (see, e.g. \cite{McWeeny,McWeeny-rev}) associated with
atoms, bonds or molecules depending on the specific type of
chemical bonding in the system.

We believe that the formalism based on group functions is the most
appropriate one for the derivation of any embedding scheme. A
rather general method based on overlapping (not strongly
orthogonal) group functions \cite{my-AD-1,my-AD-2} is presently
being developed in our laboratory. Our method which is similar in
spirit to some one-electron methods
\cite{Shidlovskaya-2002,Fornili-Sironi-Raimondi-2003,Mo-Gao-2000}
is based on construction of strongly localized orbitals which are
designed to represent the true electronic density of the entire
system via a combination of elementary densities associated in
simple cases with atoms, ions and/or bonds. Our initial effort in
this project is focused on the development of an embedding scheme
based on the HF approximation and applied to point defects in the
bulk or at surfaces of periodic crystals. Our intention is to
create a rather general technique which can be valid for systems
of different chemical character, ranging from purely ionic to
strongly covalent (note that our method cannot be directly applied
to metals). Therefore, the proper choice of the localization
technique which can deliver localized orbitals across a wide range
of systems with various character of chemical bonding is crucial
for our method to work for those systems.

It is the main objective of the present paper to critically
analyze and develop further a number of localization methods in
order to verify their ability to describe a wide range of
different chemical bondings in \emph{periodic crystals}. Two
systems are considered in the present paper in detail, MgO and Si
bulk crystals, which are examples of extreme ionic and covalent
bonding, respectively. Note, however, that the method we suggest
is not limited to periodic systems and, with some insignificant
modifications, can also be applied e.g. to infinite amorphous and
finite biological systems. The application of the present method
to those systems will be a matter of future publications.

It is relevant to mention, as far as the localization methods are
concerned, that there are several methods developed
\cite{Marzari97,Dovesi-Saunders2001} for obtaining
\emph{orthogonal} localized orbitals (i.e. Wannier functions) out
of the Bloch-like solutions of the HF or Kohn-Sham (KS) equations
\cite{Parr-Yang} for periodic crystals. Due to the built-in
orthogonality even strongly localized Wannier functions have
long-range tails which make these functions nontransferable to
other systems, e.g. when a chemical bond between the same species
is placed in a different chemical environment. That is why our
interest is focused on construction of \emph{non-orthogonal
localized orbitals} which do not have this disadvantage and thus
are more appropriate for our purposes.

The plan of the paper is the following. Our philosophy in
constructing localized orbitals as well as a short overview of
existing localization methods  is given in section
\ref{sec:Localisation-methods} with special emphasis on the
methods used in our present work. All the necessary notations are
also introduced there. In section 3 we describe our implementation
of some of the methods and their application to MgO and Si
crystals. The paper is finished with a short discussion and
conclusions in section 4.

\section{Localization methods\label{sec:Localisation-methods}}

\subsection{General philosophy}

In order to describe the crystal as a set of overlapping localized
functions, $\left\{ \widetilde{\varphi}_{a}(\mathbf{r)}\right\} $,
which are given as a linear combination of the original canonical
set $\left\{ \varphi_{i}^{c}(\mathbf{r)}\right\} $ (and which thus
span the same occupied Fock space), one has first to identify the
regions of space where each of the functions
$\widetilde{\varphi}_{a}(\mathbf{r)}$ have to be localized. This
question can be viewed as purely technical since, any linear
combination of the canonical set will give the same electron
density $\rho(\mathbf{r)}$. We, however, adopt in our work a
strategy based on the chemistry of the system in question. Namely,
the choice of the localization regions is based on the type of the
chemical bonding, e.g. on atoms in the cases of atomic or ionic
systems, on two atoms in the case of covalent bonding, etc. A more
complicated choice may be necessary in the cases of intermediate
bonding. Several different nonequivalent regions may be necessary
to represent a crystal unit cell which can then be periodically
translated to reproduce the whole infinite crystal. Note that
there could be several localized orbitals associated with every
such a region forming together an electronic group
\cite{McWeeny,EMC-1}. For instance, in the case of the Si crystal
one needs four localized regions associated with four bonds; each
bond is represented by a single double occupied localized orbital.

Once the occupied Fock space is obtained via a set of canonical
orbitals and localized regions are identified, it is necessary to
find such linear combination of canonical orbitals which are
localized in each of the regions. The topic of construction of
localized (non-canonical) molecular orbitals (MO) out of
delocalized canonical solutions of the HF or Kohn-Sham equations
is an old one \cite{Loc-Deloc} and many methods have since been
developed.

Let us assume that a canonical solution of the restricted HF
equations for the entire system (a closed shell crystal) is known
\cite{McWeeny}:\begin{equation}
\widehat{F}\varphi_{i}^{c}(\mathbf{r})=\varepsilon_{i}\varphi_{i}^{c}(\mathbf{r})\label{eq:HF-eq}\end{equation}
\begin{equation}
\varphi_{i}^{c}(\mathbf{r})=\sum_{\mu}C_{\mu
i}^{c}\chi_{\mu}(\mathbf{r})\label{eq:CO_via_AO}\end{equation}
where $\widehat{F}$ is the Fock operator,
$\varphi_{i}^{c}(\mathbf{r})$ is a spin-independent canonical MO
(CMO) which is expanded over a set of atomic orbitals (AOs)
$\chi_{\mu}(\mathbf{r})$. The electronic density of the system
\begin{equation}
\rho(\mathbf{r})=2\sum_{i}^{occ}\left|\varphi_{i}^{c}(\mathbf{r})\right|^{2}\label{eq:density-orth}\end{equation}
contains the summation only over occupied CMOs thus ensuring the
correct normalization to the number $N$ of the electrons in the
system. If an arbitrary (generally \emph{non-unitary})
transformation $\mathbf{U=\parallel}U_{aj}\mathbf{\parallel}$ of
the CMOs within the occupied subspace is performed,
\begin{equation}
\widetilde{\varphi}_{a}(\mathbf{r})=\sum_{j}^{occ}U_{aj}\varphi_{j}^{c}(\mathbf{r})\equiv\sum_{\mu}\widetilde{C}_{\mu
a}\chi_{\mu}(\mathbf{r})\label{eq:U-transf}\end{equation} than the
expression for the density via the new set of orbitals should
contain the inverse of the overlap matrix
$\mathbf{\widetilde{S}=\parallel}\widetilde{S}_{ab}\mathbf{\parallel}$
\cite{McWeeny}: \begin{equation}
\rho(\mathbf{r})=2\sum_{ab}^{occ}\widetilde{\varphi}_{a}(\mathbf{r})\left(\widetilde{S}\right)_{ab}^{-1}\widetilde{\varphi}_{b}^{*}(\mathbf{r})\label{eq:density-non-orth}\end{equation}
where $\widetilde{S}_{ab}=\left\langle
\widetilde{\varphi}_{a}(\mathbf{r})\right|\left.\widetilde{\varphi}_{b}(\mathbf{r})\right\rangle
$ is the overlap integral. Note that the two representations of
the electron density, Eqs. (\ref{eq:density-orth}) and
(\ref{eq:density-non-orth}) are absolutely equivalent. Moreover,
any linear combination (\ref{eq:U-transf}) of the occupied CMOs
leads to the same density. If the transformation is unitary, then
the overlap matrix is a unity matrix and the density takes on its
{}``diagonal{}`` form (\ref{eq:density-orth}).

In general, any localization procedure is equivalent to some
transformation $\mathbf{U}$ of the CMOs. Suppose, we would like to
obtain $n$ localized MOs (LMOs) in some region $A$. To find the
necessary transformation, one can formulate an optimization
(minimization or maximization) problem for some specific
\emph{localizing functional} $\widetilde{\Omega}_{A}\left[\left\{
\widetilde{\varphi}_{a}\right\} \right]$ with the constraint that
the LMOs associated with region $A$ are orthonormal (of course,
LMOs associated with different regions will not be orthogonal in
general). We shall limit ourselves with such functionals which are
invariant under arbitrary unitary transformations of LMOs, i.e.
which in fact depend on the orbitals $\left\{
\widetilde{\varphi}_{a}\right\} $ via invariants in the form of
the non-diagonal {}``density''\begin{equation}
\sigma_{A}(\mathbf{r},\mathbf{r}^{\prime})=\sum_{a=1}^{n}\widetilde{\varphi}_{a}(\mathbf{r})\widetilde{\varphi}_{a}^{*}(\mathbf{r}^{\prime})\label{eq:region-A_density}\end{equation}
 constructed out of the LMOs associated with region $A$, i.e. $\widetilde{\Omega}_{A}\left[\left\{ \widetilde{\varphi}_{a}\right\} \right]\equiv\Omega_{A}\left[\sigma_{A}\right]$.
We shall see in a moment that his requirement ensures an existence
of a simple eigenvalue-like problem for the LMOs. Note in passing
that some other types of functionals are also sometimes used which
do not fall within this category. For instance, Admiston and
Ruedenberg proposed to find the maximum of the self-repulsion
energy \cite{Edmiston-Ruenberg}, while later on von Niessen
suggested to maximize the charge density overlap functional
\cite{vonNiessen71}. Since the mentioned functionals are not
invariant under unitary transformations of LMOs and are also quite
expensive computationally, we do not consider them in the
following. The quantity
$\sigma_{A}(\mathbf{r},\mathbf{r}^{\prime})$ will be referred to
in the following as the \emph{region electron density} or region
density for short.

To obtain all $n$ LMOs associated with region $A$, an optimum of
the following functional is sought for: \begin{equation}
\Omega_{A}^{\prime}\left[\sigma_{A}\right]=\Omega_{A}\left[\sigma_{A}\right]-\sum_{a,b=1}^{n}\xi_{ab}\left(\left\langle
\widetilde{\varphi}_{a}\right|\left.\widetilde{\varphi}_{b}\right\rangle
-\delta_{ab}\right)\label{eq:functional-to-opt}\end{equation}
where $\xi_{ab}$ are the corresponding Lagrangian multipliers.
Because the actual dependence of the functional
$\Omega_{A}\left[\sigma_{A}\right]$ on the orbitals is built-in
via the region density (\ref{eq:region-A_density}), the functional
derivative
$\frac{\delta\Omega_{A}}{\delta\widetilde{\varphi}_{a}^{*}(\mathbf{r)}}$
can always be written using an operator
$\widehat{\Omega}_{A}(\mathbf{r)}$ defined through an identity
$\frac{\delta\Omega_{A}}{\delta\widetilde{\varphi}_{a}^{*}(\mathbf{x)}}=\widehat{\Omega}_{A}\widetilde{\varphi}_{a}(\mathbf{x)}$
since
$\frac{\delta\sigma_{A}(\mathbf{r,}\mathbf{r^{\prime}})}{\delta\widetilde{\varphi}_{a}^{*}(\mathbf{x)}}=\widetilde{\varphi}_{a}(\mathbf{r)\delta(x}-\mathbf{r^{\prime})}$.
Examples illustrating this point will be given below. We shall
refer to the operator $\widehat{\Omega}_{A}(\mathbf{r)}$ as the
\emph{localization operator} in the following. An important
property of the localization operator is that it can also be
considered as a functional of the region density
(\ref{eq:region-A_density}), i.e. it preserves the invariance
property of the localizing functional it is built from.

Using standard methods, i.e. setting the variational derivative of
the functional (\ref{eq:functional-to-opt}) with respect to the
orbital $\widetilde{\varphi}_{a}^{*}(\mathbf{r)}$ to zero and then
performing a unitary transformation of the LMOs which diagonalizes
the matrix of Lagrangian multipliers, one can easily obtain the
following equations for the LMOs sought for:\begin{equation}
\widehat{\Omega}_{A}\widetilde{\varphi}_{a}(\mathbf{r)}=\lambda_{a}\widetilde{\varphi}_{a}(\mathbf{r})\label{eq:eigenv-problem-operator}\end{equation}
or, in the matrix form, \begin{equation}
\sum_{j}^{occ}\Omega_{ij}^{A}U_{aj}=\lambda_{a}U_{ai}\label{eq:eigenv-problem-matrix}\end{equation}
where the matrix $\Omega_{ij}^{A}$ is given via matrix elements of
the operator $\widehat{\Omega}_{A}$ calculated using canonical
orbitals $\varphi_{i}^{c}(\mathbf{r)}$ and
$\varphi_{j}^{c}(\mathbf{r)}$.

Equations (\ref{eq:eigenv-problem-operator}) resemble an
eigenvalue problem for the operator $\widehat{\Omega}_{A}$. Note,
however, that in some cases the localization operator may still
depend on the region density and thus on the orbitals themselves.
Therefore, similarly to the HF or Kohn-Sham problem, the system of
equations (\ref{eq:eigenv-problem-operator}) should be solved
self-consistently.

The eigenvalue problem (\ref{eq:eigenv-problem-operator}) or
(\ref{eq:eigenv-problem-matrix}) may give a set of solutions from
which only the first (in the case when $\Omega_{A}$ is minimized)
or the last (maximized) $n$ solutions should be chosen. If the
localization criterion (i.e. the functional $\Omega_{A}$) used is
appropriate, then (i) the chosen $n$ solutions would have close
eigenvalues $\lambda_{a}$ which corresponds to their similar
localization in region $A$, and (ii) the gap in the eigenvalues
$\lambda_{a}$ between the chosen $n$ and other solutions is
considerable, i.e. the other solutions have much worse
localization in region $A$ (cf. \cite{Whitt-Pakk}). By collecting
LMOs from all regions in the unit cell and then translating those
over the whole crystal it should be possible to span the whole
occupied Fock space and thus construct the total electron density
(\ref{eq:density-non-orth}).

One point is in order now. So far we have assumed that the set of
canonical MOs which span the occupied part of the Fock space is
already known. In other words, the procedure consists of two
steps: firstly, a HF (or Kohn-Sham) problem is solved and thus the
occupied Fock space is determined, and, secondly, the LMOs are
obtained by finding an appropriate linear combinations of the
canonical orbitals within this space. However, it is also possible
to formulate the problem in such a way that LMOs are obtained
together with the set of canonical orbitals in a single step
\cite{Loc-Deloc}. In this method a localization criterion is
considered alongside the energy minimization leading to a set of
so-called Adams-Gilbert (AG) equations (see, e.g. \cite{Gilbert})
which are solved in a self-consistent manner. For instance, a
projection operator on the subspace of the LMOs was used by Stoll
\emph{et. al.} \cite{Stoll1980} as the specific localization
method. This technique was implemented in \cite{Shidlovskaya-2002}
for the embedded molecular cluster calculations. The LMOs
resulting from a single AG calculation are orthogonal as solutions
of a single secular problem. The first eigenvectors obtained will
show strong localization within the chosen region $A$; other
eigenvectors will be much less localized and can usually be
distinguished by a gap in their eigenvalues as explained above. To
obtain LMOs strongly localized in a different region $A^{\prime}$,
one has to solve the AG equations once again using another
localization criterion and then pick up the necessary number of
the most localized orbitals. Repeating this procedure across the
entire system, the whole occupied Fock space can be split into
sets of mutually non-orthogonal LMOs. Of course, in the case of
the perfect crystals this procedure should only be applied to
various localization regions within the primitive unit cell owing
to crystals periodic symmetry.

Note that it is also possible to obtain all the LMOs corresponding
to several localization regions at once within the same
self-consistent calculation by solving the necessary sets of
eigenproblems associated with each region
\cite{Fornili-Sironi-Raimondi-2003,Mo-Gao-2000}. The LMOs obtained
using this technique are known as extremely localized MOs. This
method is quite expensive computationally since the overlap
between LMOs localized in different regions in space changes in
the course of the iteration procedure and this affects the
convergence. However, as will be discussed in the following, this
technique is a logical step forward in the future development of
our method to be presented in section 3.

There are many ways in which the localizing functional
$\Omega_{A}$ can be chosen. Some of these methods which will be
utilized in the present work will be considered below in more
detail.

\subsection{Methods based on functionals linear in region density}

In a number of methods \cite{Mayer1996} the localizing functional
is proportional to the non-diagonal density
(\ref{eq:region-A_density}) and thus can be represented as a
Hermitian bilinear functional with respect to the LMOs of the
following general form:\begin{equation}
\Omega_{A}=\int\left[\widehat{\Omega}_{A}\sigma_{A}(\mathbf{r},\mathbf{r}^{\prime})\right]_{\mathbf{r^{\prime}\rightarrow
r}}d\mathbf{r}=\sum_{a=1}^{n}\int\widetilde{\varphi}_{a}^{*}(\mathbf{r})\widehat{\Omega}_{A}\widetilde{\varphi}_{a}(\mathbf{r})d\mathbf{r}\equiv\sum_{a=1}^{n}\sum_{jk}^{occ}U_{aj}^{*}\Omega_{jk}^{A}U_{ak}\label{eq:F_A_first_group}\end{equation}
where $\widehat{\Omega}_{A}$ is some localization operator and the
Hermitian matrix
$\mathbf{\Omega}^{A}=\parallel\Omega_{jk}^{A}\parallel$ can easily
be written in terms of the canonical MOs using the definition
(\ref{eq:CO_via_AO}):

\begin{equation}
\Omega_{jk}^{A}=\left\langle
\varphi_{j}^{c}\right|\widehat{\Omega}_{A}\left|\varphi_{k}^{c}\right\rangle
=\sum_{\mu,\nu}C_{\mu j}^{c*}C_{\nu k}^{c}\left\langle
\chi_{\mu}\right|\widehat{\Omega}_{A}\left|\chi_{\nu}\right\rangle
\label{eq:Q-matrix-1st-group}\end{equation} For all methods of
this group both the operator $\widehat{\Omega}_{A}$ and the matrix
$\mathbf{\Omega}^{A}$ do not depend on the LMOs sought for so that
in order to obtain the localized orbitals one has simply to find
the eigenvectors of the matrix $\mathbf{\Omega}^{A}$ using Eq.
(\ref{eq:eigenv-problem-matrix}). Two methods of this group are
implemented in our work and will be considered in the following in
more detail.

\subsubsection*{Mulliken's net population (method M)}

Magnasco-Perico criterion maximizes Mulliken's \cite{Mulliken} net
atomic population produced by the LMOs in the selected region
\cite{Magnasco-Perico,Mayer1996}. In this case the matrix
$\mathbf{\Omega}^{A}$ is chosen in the following
form:\begin{equation} \Omega_{jk}^{A}=\sum_{\mu,\nu\in A}C_{\mu
j}^{c*}S_{\mu\nu}C_{\nu
k}^{c}\label{eq:Q-matrix_for_Mulliken}\end{equation} where
$S_{\mu\nu}$ is the overlap integral between two AOs $\chi_{\mu}$
and $\chi_{\nu}$. The summation here is performed over AOs which
are centered in the chosen region $A$. Thus, in practice the
localization region in this method is specified by a selection of
AOs in Eq. (\ref{eq:Q-matrix_for_Mulliken}). This way one can make
the LMOs to have the maximum contribution from the specified AOs
in region $A$. Sometimes a different choice of AOs may lead to
physically identical localization (see the next section). This
method will be referred to as method M.

\subsubsection*{The projection on the atomic subspace (method P)}

The Roby's population maximization \cite{Roby74} gives LMOs for
which the projection on the subspace spanned by the basis orbitals
centered within the selected region is a maximum, or is at least
stationary \cite{Mayer1996}. In this method the localization
operator $\widehat{\Omega}_{A}$ in Eq. (\ref{eq:F_A_first_group})
is chosen in the form of a projection operator:

\begin{equation}
\widehat{\Omega}_{A}=\sum_{\mu,\nu\in
A}\left|\chi_{\mu}\right\rangle
(\mathbf{S}_{A}^{-1})_{\mu\nu}\left\langle
\chi_{\nu}\right|\label{eq:F-operator_for_Oper_method}\end{equation}
where $\mathbf{S}_{A}^{-1}$ stands for the inverse of the overlap
matrix $\mathbf{S}_{A}$ defined on all AOs $\mu,\nu\in A$. Note
that operator $\widehat{\Omega}_{A}$ is idempotent:
$\left(\widehat{\Omega}_{A}\right)^{2}=\widehat{\Omega}_{A}$. It
projects any orbital into a subspace spanned by the AOs associated
with region $A$ only. In particular,
$\widehat{\Omega}_{A}\left|\chi_{\mu}\right\rangle
=\left|\chi_{\mu}\right\rangle $. The detailed expression for the
matrix $\mathbf{\Omega}^{A}$ is then:

\begin{equation}
\Omega_{jk}^{A}=\sum_{\lambda,\tau}C_{\lambda j}^{c*}C_{\tau
k}^{c}\left[\sum_{\mu,\nu\in
A}S_{\lambda\mu}(\mathbf{S}_{A}^{-1})_{\mu\nu}S_{\nu\tau}\right]\label{eq:Q-matrix_Oper_method}\end{equation}
Here the first double summation is performed over all AOs of the
system. Region $A$ is also defined via a subset of AOs: by
choosing particular AOs one ensures the maximum overlap of the
LMOs with them. It is seen that this method, which will be
referred to as method P, although different in the implementation,
is very similar in spirit to the previous method M.

\subsubsection*{Other methods}

Note that several other methods \cite{Mayer1996} also belong to
this class of methods. Since we are not using them here, we shall
only mention some of them. Bader's method (see also \cite{Bader})
is computationally expensive and leads to LMOs with
discontinuities at the boarder of the localization regions. The
widely used Pipek-Mezey localization scheme \cite{Pipek-Mezey}
could be described as the maximization of the Mulliken's gross
atomic population. The Pipek-Mezey functional corresponds also to
a minimization of the number of atoms over which the LMO is to be
spread. This method is very similar to the Mulliken's net
population method considered above, although is slightly more
computationally expensive. All population methods have an
advantage of being very simple in the implementation which results
in fast non-self-consistent algorithms: indeed, only overlap
integrals are to be computed. Note that instead of the overlap,
one can also maximize an exchange interaction of the LMOs with a
set of AOs in region $A$ \cite{Whitt-Pakk}.

Perhaps the most widely used, due to its relatively low
computational cost, is the Foster-Boys \cite{FosterBoys} method in
which the dipole moment matrix element between so-called exclusive
orbitals is maximized. The efficiency of the HF method was
improved in \cite{Anikin2003} \textbf{}by using localized orbitals
constructed from the Foster-Boys method as AOs\textbf{.} Recently
\cite{Dovesi-Saunders2001} the Wannier functions were calculated
for periodic Si and MgO crystals using the modified version of the
Foster-Boys method which ensured better localization within the
cell volume. Note that, according to Ref. \cite{Mayer1996}, the
Pipek-Mezey functional, unlike the Fosters-Boys method, preserves
the $\sigma/\pi$ separation of double bonds, which is in chemistry
usually preferred over the $\tau$ picture (where the orbitals are
proportional to the linear combinations $\sigma+\pi$ and
$\sigma-\pi$) associated with the Fosters-Boys localization
procedure.

\subsection{Methods based on functionals bilinear in region density\label{sub:bilinear_functionals}}

More complicated localization procedures can be constructed if the
localizing functional is bilinear in non-diagonal density
(\ref{eq:region-A_density}) of region $A$ in question, i.e. is of
the fourth order with respect to the LMOs sought for. Three
important general points should be mentioned: (i) for all methods
of this group the localizing operator $\widehat{\Omega}_{A}$ is
linear in the density $\sigma_{A}(\mathbf{r},\mathbf{r}^{\prime})$
and (ii) is thus invariant under any unitary transformation of the
LMOs; (iii) therefore, one still has the secular problem
(\ref{eq:eigenv-problem-operator}) or
(\ref{eq:eigenv-problem-matrix}) for the LMOs in this case;
however, it is to be solved \emph{self-consistently}.

\subsubsection*{Minimization of the HF energy of a structure element (method E)}

In this paper we shall only apply one method of this group in
which the functional $\Omega_{A}$ is chosen as the HF energy of a
finite fragment (structure element (SE), cf. \cite{EMC-1}) of the
system. This method, which will be referred to as method E in the
following, in its simplest version of the localization on a single
atom originates from Adams \cite{Adams}. This method was recently
used in \cite{Abarenkov2004} to derive an embedding potential
provided by a part of a molecule. Method E is based on an
intuitive idea that every \emph{finite} system, e.g. a SE, will
try to find an energetically favorable ground state which will be
localized in space. The SE could be an atom, group of atoms or a
bond. In the latter case the SE for Si represents essentially a
hydrogen-like molecule consisting of fragments of two nearest
atomic cores each of charge $+e$ ($e$ is the electron charge) and
two electrons of opposite spins (see the next section).

The eigenvalue problem (\ref{eq:eigenv-problem-matrix}) in this
case is nothing but the usual Hartree-Fock-Roothaan (HFR) problem
\cite{McWeeny} for molecular orbitals of region $A$, i.e. the
elements of the matrix $\mathbf{\Omega}^{A}$ are matrix elements
\[ \Omega_{jk}^{A}=\left\langle
\varphi_{j}^{c}\right|\widehat{\Omega}_{A}\left|\varphi_{k}^{c}\right\rangle
\] where $\widehat{\Omega}_{A}\equiv\widehat{F}_{A}$ is the usual
HF operator of the SE containing both electron-electron and
electron-core interactions. The peculiarity of this case is that
the MOs (which are, in fact, LMOs) are expanded not via AOs but
rather via canonical HF orbitals for the whole system which, in
turn, are linear combinations of AOs of the whole system as in Eq.
(\ref{eq:CO_via_AO}). In other words, the difference with the
usual setup of the HFR problem is that in our case the preset
linear combinations (\ref{eq:CO_via_AO}) of the AOs of the whole
system are used in place of the AOs themselves. Correspondingly,
this method is computationally expensive since all the one- and
two-electron integrals which are necessary for the construction of
the Fock matrix $\mathbf{\Omega}^{A}$ are expressed as double and
quadruple sums of the corresponding AO integrals.

\section{Localized molecular orbitals for Si and MgO bulk crystals}

In this section we shall examine LMOs obtained using three methods
described above. Two crystalline systems with extreme types of
chemical binding will be considered: MgO (ionic) and Si
(covalent).

\subsection{General method}

We are interested in calculating LMOs for perfect periodic solids.
In this case the canonical MOs are Bloch-like solutions of the HF
or Kohn-Sham equations. This means that the eigenvalue problem
(\ref{eq:eigenv-problem-matrix}) is of an infinite dimension which
make the calculation quite complicated; in addition, it would be
necessary to use (and modify) a periodic electronic structure
code. To avoid these difficulties, we suggest a very simple
procedure based on a cluster method. The basic idea relies on the
fact that when the cluster size is increased, the distribution of
the electron density in its central region should become closer to
the actual electron distribution of the infinite periodic system.

Our method is based on the following steps:

\begin{enumerate}
\item analyze the known electron charge density $\rho(\mathbf{r)}$
of the 3D periodic system in question to identify regions $A$,
$B$, $C$, etc. which can each be associated with even numbers of
localized electrons, e.g. atoms, ions, bonds; the density
$\rho(\mathbf{r)}$ can be known from literature or own
calculations; \item consider a quantum cluster which contains
region $A$ in its center (or close to it); terminate the cluster
using pseudoatoms (see below) and/or an array of point charges to
reproduce the correct Madelung field; obtain the occupied
canonical orbitals $\varphi_{i}^{c}(\mathbf{r})$ for the whole
cluster; \item then consider a localization problem for region $A$
using one of the methods of the previous section; this should give
the necessary number of LMOs $\widetilde{\varphi}_{a}(\mathbf{r})$
($a=1,\ldots,n$) as a linear combination of the occupied canonical
MOs $\left\{ \varphi_{i}^{c}(\mathbf{r})\right\} $, Eq.
(\ref{eq:U-transf}); note that in some cases when e.g. pseudoatoms
are used to terminate the cluster, their contribution to the LMOs
should be removed and the orbitals renormalized; \item repeat the
procedure 2-3 for larger clusters to ensure that the LMOs obtained
have converged; \item if other types of regions exist, repeat
steps 2-4 for those regions as well; when finished, LMOs for the
whole unit cell should be available; sometimes (as is the case for
silicon), LMOs of some other regions can be obtained without
additional calculation by simply translating and possibly rotating
the LMOs of a single region; \item the LMOs within the primitive
cell can be displaced by all possible lattice translations to
obtain the complete set of crystal LMOs spanning the complete
occupied Fock space; these can now be employed for the calculation
of the density $\widetilde{\rho}(\mathbf{r})$ according to Eq.
(\ref{eq:density-non-orth}); note that
$\widetilde{\rho}(\mathbf{r})$ and the actual periodic density
$\rho(\mathbf{r})$, calculated using a periodic code and identical
basis set, will not be exactly the same due to a number of
approximations employed here in calculating the LMOs (see below).
\end{enumerate}
Some general comments of our general method are necessary at this
point. Firstly, the convergence of our procedure with the cluster
size depends on the {}``boundary conditions'' used in every case,
i.e. it depends on the way the cluster is terminated; it is
nothing but the embedding method itself. The latter is, however,
unknown. Therefore, our procedure can be considered only as the
first iteration and thus larger cluster sizes are expected. In
principle, when the LMOs are obtained, one can use them for a new
set of embedding calculations to obtain a better approximation for
them, etc. In this case smaller cluster sizes may only be
necessary.

Secondly, it has already been mentioned that any linear
combination of the canonical set of MOs should lead to the same
electron density. Therefore, one may think that the density
$\widetilde{\rho}(\mathbf{r})$ obtained using the procedure
outlined above will always result in the correct electron density
$\rho(\mathbf{r})$. This is, however, do not need to be the case
due to a number of approximations adopted. Indeed, we only
consider finite clusters with \emph{ad-hoc} boundary conditions;
in addition, the contribution of boundary cluster atoms may be
modified when pseudoatoms are used. Finally, the cluster size may
be insufficient to accommodate completely the LMOs. Therefore, the
obtained occupied Fock space will never be exactly the same as
that obtained using the periodic calculation. Hence, the
comparison of the electron densities
$\widetilde{\rho}(\mathbf{r})$ and $\rho(\mathbf{r})$ may indicate
on the quality of the calculated LMOs, and this method will be
used in this paper.

The calculation of $\widetilde{\rho}(\mathbf{r})$ is performed by
exploiting the periodic symmetry and representing the LMOs as an
integral over the Brillouin zone. This method is exact and does
not depend on the degree of localization of the LMOs. It also
allows exact handling of the inverse of the overlap matrix in Eq.
(\ref{eq:density-non-orth}). Details of the method will be
published elsewhere. To calculate the reference density
$\rho(\mathbf{r})$, we used our cluster calculations in the
following way: (i) a parallelepiped in the central part of the
cluster with the sides along the primitive lattice translations
$\mathbf{a}_{1}$, $\mathbf{a}_{2}$ and $\mathbf{a}_{3}$ which is
equivalent to the primitive unit cell is identified; its density
is denoted $\rho_{c}(\mathbf{r})$; (ii) the density of the whole
3D crystal is then modelled as
$\rho(\mathbf{r)}\equiv\rho_{c}(\mathbf{r}_{1})$, where
$\mathbf{r}_{1}$ is obtained from $\mathbf{r}$ by removing any
lattice translations (this is most conveniently performed by first
calculating fractional coordinates of $\mathbf{r}$ in terms of
$\mathbf{a}_{1}$, $\mathbf{a}_{2}$ and $\mathbf{a}_{3}$ and then
removing integral parts from them). The larger size of the cluster
is used in the calculations, the better approximation for the
density $\rho(\mathbf{r})$ will be obtained in this way.

All numerical calculations reported in this paper were done using
the HF method and the Gamess-UK \cite{GAMESS-UK} code within the
pseudopotential approximation.

\subsection{Localized orbitals for the MgO bulk}

\begin{figure}[h]
\begin{center}\includegraphics[%
  height=7cm,
  keepaspectratio]{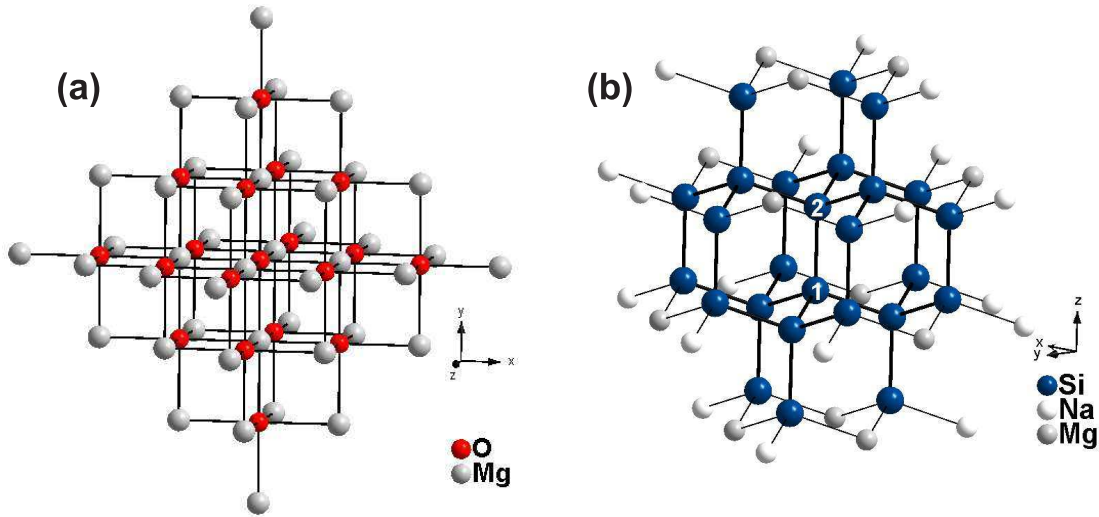}\end{center}

\caption{The largest quantum clusters used in our HF calculations
to model (a) MgO and (b) Si crystals. Points charges surrounding
the MgO cluster to simulate the Madelung field are not shown. One
electron Na and two electron Mg pseudoatoms (shown) were used in
(b) to saturate bonds with boundary Si atoms (see text).
\label{Fig:MgO+Si_large}}
\end{figure}

A sequence of three finite clusters of increased size Mg$_{6}$O,
Mg$_{38}$O$_{13}$ and Mg$_{44}$O$_{19}$ containing 7, 51 and 63
\textbf{}atoms, respectively, and surrounded by an array of nearly
10$^{3}$ point charges of $\pm2e$ to simulate the Madelung field
was considered. The largest cluster used is shown in Fig.
\ref{Fig:MgO+Si_large}
(\emph{a}).%
To consider explicitly only the valence electrons, for both Mg and
O we used coreless Hartree-Fock pseudopotentials (CHF) with LP-31G
basis set from Ref. \cite{CHF:Melius-Goddard74}. The number of
electrons in each cluster was calculated by adopting a well known
ionic character of the MgO crystal, i.e. by assuming that every
Mg$^{2+}$ ion is associated with no electrons, while every
O$^{2-}$ ion has eight electrons. The LP-31G basis set was used in
all our calculations and the distance between the nearest Mg and O
atoms is 2.122 $\textrm{Å}$ \cite{irreg-MgO}.

After the HF solution was obtained for every cluster, we applied
the three localization procedures (methods M, P and E) considered
in the previous section to obtain the LMOs for this system. Since
there are only two atoms in the primitive cell, Mg and O, and it
is well known that the valence electron density is localized
predominantly on the O atoms, one can choose essentially a single
region $A$ within the primitive unit cell to localize the LMOs
into, namely, on the O atom. We should expect four LMOs localized
on every O atom: one of the $s$ type,
$\widetilde{\varphi}_{s}(\mathbf{r})$, and three of the $p$ type,
$\widetilde{\varphi}_{p_{x}}(\mathbf{r})$,
$\widetilde{\varphi}_{p_{y}}(\mathbf{r})$ and
$\widetilde{\varphi}_{p_{z}}(\mathbf{r})$.

Therefore, when applying the methods M and P we used the $s$ and
$p_{x}$, $p_{y}$ and $p_{z}$ type AOs centered on the O atom in
the center of every cluster when applying Eqs.
(\ref{eq:Q-matrix_for_Mulliken}) and
(\ref{eq:F-operator_for_Oper_method}). In the case of method E, we
considered the HF problem for a single oxygen ion O$^{2-}$ in the
basis set of all occupied canonical MOs of the entire cluster. In
every case exactly four LMOs were obtained as having the smallest
eigenvalues; other states were found to be separated by a
considerable gap.

\begin{figure}
\begin{center}\includegraphics[%
  height=7cm,
  keepaspectratio]{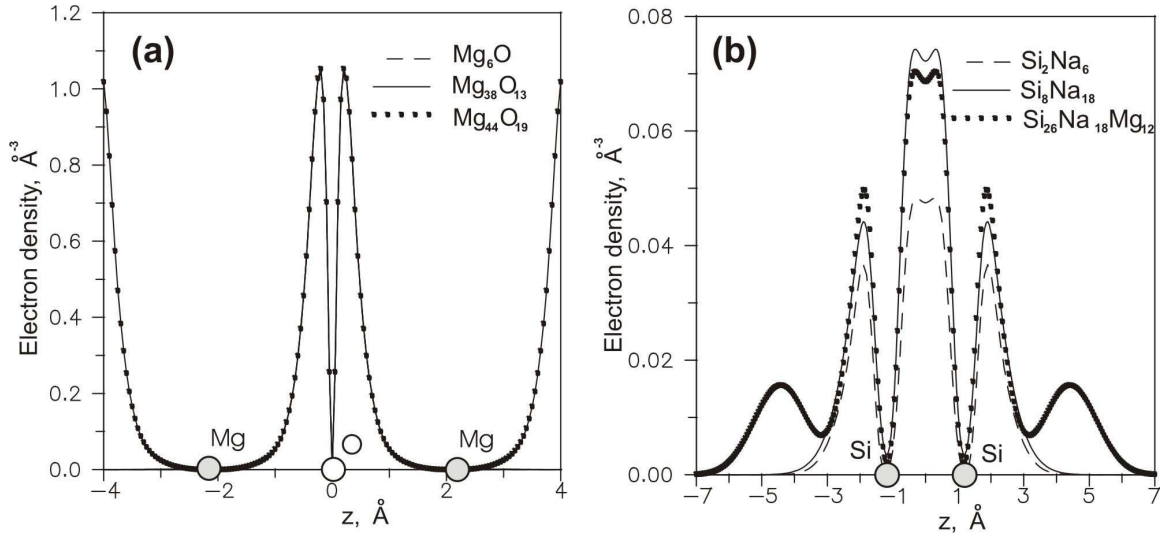}\end{center}

\caption{The HF electron density for (a) MgO and (b) Si for the
three clusters studied in each case: (a) along the line passing
through the central O and the nearest Mg atoms; (b) across the
central Si-Si bond. Positions of Mg, O and Si atoms are indicated.
\label{Fig:MgO-cluster-Ro} }
\end{figure}

The HF electron density through the central O atom for every
cluster
is shown in Fig. \ref{Fig:MgO-cluster-Ro} (a). %
 One can see that the density is perfectly converged already for the
smallest of the \textbf{}clusters. This means that the electron
density in the center of any of the clusters can be considered as
being very close to the density of the actual 3D periodic crystal
calculated in the HF approximation using the same basis set.

\begin{figure}
\begin{center}\includegraphics[%
  height=7cm,
  keepaspectratio]{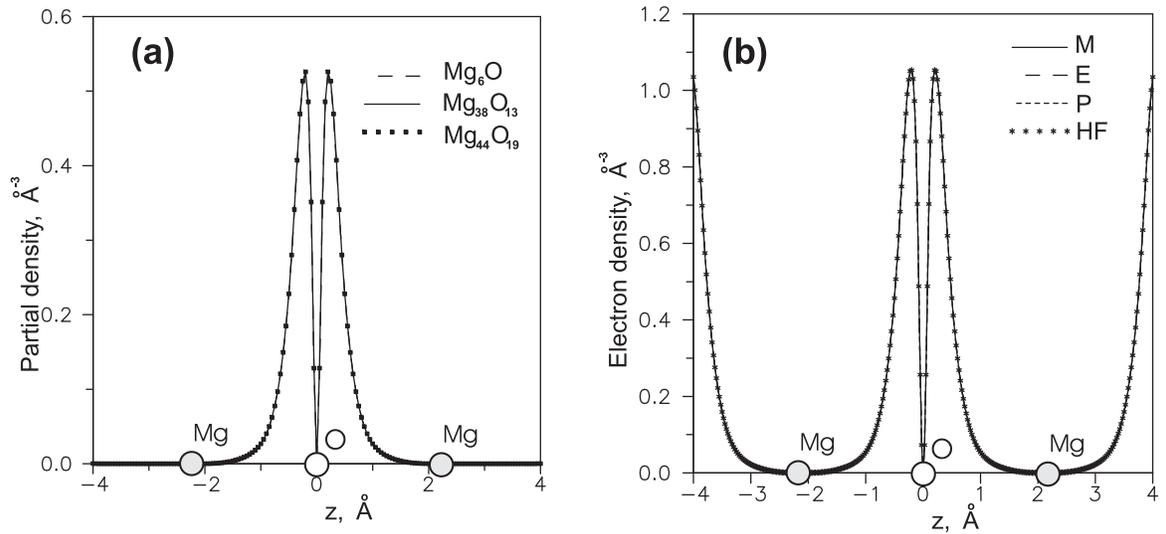}\end{center}

\caption{(a) Partial density $\rho_{O}(\mathbf{r})$, Eq.
(\ref{eq:O-partial-density}), calculated using method M for all
clusters; (b) electron densities of the MgO crystal constructed
from LMOs obtained using methods M, P and E (lines) are compared
with the HF density calculated from the middle of the largest
cluster (stars). All densities are shown along the Mg-O-Mg
direction. Mg and O atoms are indicated on the picture.
\label{Fig:LMOs-for-MgO}}
\end{figure}

The partial oxygen electron density \begin{equation}
\rho_{O}(\mathbf{r})=\widetilde{\varphi}_{s}^{2}(\mathbf{r})+\widetilde{\varphi}_{p_{x}}^{2}(\mathbf{r})+\widetilde{\varphi}_{p_{y}}^{2}(\mathbf{r})+\widetilde{\varphi}_{p_{z}}^{2}(\mathbf{r})\label{eq:O-partial-density}\end{equation}
can be conveniently used to characterize the localization of the
obtained LMOs $\widetilde{\varphi}_{s}(\mathbf{r})$,
$\widetilde{\varphi}_{p_{x}}(\mathbf{r})$, etc. We compare
$\rho_{O}(\mathbf{r})$ obtained for all clusters using method M in
Fig. \ref{Fig:LMOs-for-MgO} (a). Note that other
methods give practically identical densities. %
 It is seen that all four LMOs are extremely well localized on the
O atom (as one would expect for such an extremely ionic system)
and converge very quickly with the cluster size. The LMOs are
essentially identical for all three methods.

Obviously, LMOs associated with any other unit cell can now be
obtained simply by moving the calculated four LMOs by the
appropriate lattice translation. We have made a careful comparison
of the total electron density $\widetilde{\rho}(\mathbf{r})$
constructed using Eq. (\ref{eq:density-non-orth}) with the density
$\rho(\mathbf{r})$ calculated using the central part of the
largest cluster. In particular, such a comparison is shown in Fig.
\ref{Fig:LMOs-for-MgO} (b) along the (001) direction across the
central O atom. One can see that either method results in the
perfect matching between the density
$\widetilde{\rho}(\mathbf{r})$ obtained using the LMOs
(indistinguishable on the plot) and the reference density,
$\rho(\mathbf{r})$. Thus, all three localization techniques work
equally well in the case of MgO and require clusters of very
moderate sizes.

\subsection{Localized orbitals for the Si bulk}

\begin{figure}[h]
\begin{center}\includegraphics[%
  scale=0.5]{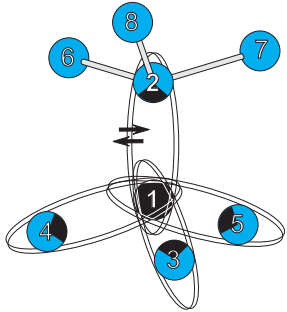}\end{center}

\caption{Four two-electron bonds associated with a primitive unit
cell in the Si crystal. \label{Fig:4-bonds}}
\end{figure}

Crystalline Si has the diamond-type lattice with the distance
between the nearest Si atoms of 2.35$\textrm{Å}$. Each Si atom is
surrounded by four neighbors forming four covalent bonds with
them. There are two Si atoms and thus eight electrons (within the
valence approximation) to be assigned to every primitive cell.
Since each bond is associated with two electrons, there should be
four bonds per cell. We expect that well localized orbitals can be
constructed for this crystal if localization regions are
associated with every two-electron bond. Therefore, in this case
we have four regions $A$, $B$, $C$ and $D$ of identical nature in
the primitive cell as shown schematically
in Fig. \ref{Fig:4-bonds}. %
Note that the choice of four inequivalent bonds is not unique. In
our choice shown in the figure all four bonds share atom 1 and can
be obtained from any single bond (e.g. the central bond between
atoms 1 and 2) by applying appropriate displacements and
rotations.

Three quantum clusters Si$_{2}$Na$_{6}$, Si$_{8}$Na$_{18}$ and
Si$_{26}$Na$_{18}$Mg$_{12}$, containing 2, 8 and 26 Si atoms were
used in our calculations; every cluster contains a single Si-Si
pair in its center as shown in Fig. \ref{Fig:MgO+Si_large} (b).
The core electrons of all Si atoms were described using the
Hay-Wadt pseudopotential \cite{lanl}. To terminate unsaturated
bonds of the Si atoms located at the boundary of the clusters, we
used Na- and Mg-like pseudoatoms which have the same
pseudopotentials as the Si atoms. The Na pseudoatoms contribute a
single electron to the cluster and are positioned at the correct
Si-Si distance to saturate a single dangling bond. Mg-like atoms
have two electrons and were used in the same way to saturate two
dangling bonds from two nearest boundary Si atoms (see Fig.
\ref{Fig:MgO+Si_large} (b)). Since the Si crystal is a highly
covalent system, the Madelung field can be considered of a
secondary importance and thus was neglected. The 66-21G basis set
\cite{66-21G} was used on the Si atoms in most cases. The basis
set on pseudoatoms Na and Mg included only $s$ type AOs.

To construct the LMOs for the Si crystal, we oriented the
coordinate system in such a way that the $z$ axis would pass along
the central Si-Si bond of every cluster. This particular choice of
the coordinate system is merely needed to simplify the choice of
the AOs to be associated with the localization region $A$. Then,
the HF solution was obtained which demonstrated a good degree of
the $sp^{3}$ hybridization, as expected. When applying the
localization methods M and P, AOs of the $s$ and $p_{z}$ types
centered on the two central Si atoms were chosen as belonging to
region $A$. In order to apply method E, the following SE was
considered in place of the central Si-Si molecule: it consisted of
two electrons and two pseudoatoms with the Si pseudopotential and
the total charge $+e$ each. Effectively, this way the SE was
chosen as a pseudo-hydrogen molecule with pseudo-hydrogen atoms at
the Si-Si distance described each by the Si pseudopotential.

By analyzing the electron density in the central region of every
cluster, we find that the largest cluster we considered is
sufficient for our purposes. As an example, we show in Fig.
\ref{Fig:MgO-cluster-Ro} (b) the electron densities across the
central Si-Si bond for the three clusters. The single LMOs
calculated for the central Si-Si bond of the three clusters using
methods M and E are shown in Fig.
\ref{Fig:M,E-cal-LMOS-comparison}. The LMOs calculated using
method P are very similar to those calculated using method M and
thus are not shown here. The important conclusion which can be
drawn from these pictures is that the LMO obtained using method M
is practically converged with the cluster size. On the contrary,
the LMO calculated using method E is not yet converged becoming
more
and more delocalized with the increase of the cluster size. %
\begin{figure}
\begin{center}\includegraphics[%
  height=7cm,
  keepaspectratio]{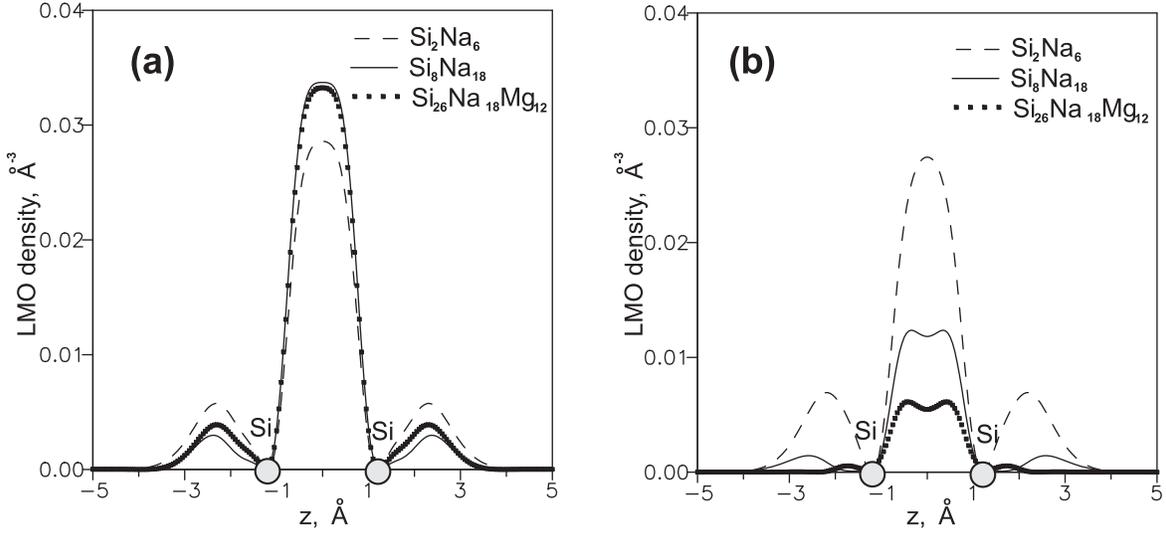}\end{center}

\caption{Square of the central bond LMO
$\widetilde{\varphi}(\mathbf{r)}$ calculated across the central
Si-Si bond for every cluster using methods (a) M and (b) E.
\label{Fig:M,E-cal-LMOS-comparison}}
\end{figure}
 Even bigger clusters are thus needed to converge the LMO using this
method. However, as was mentioned earlier in section
\ref{sub:bilinear_functionals}, larger clusters require a very
expensive procedure of calculating two-electron integrals, and
thus we did not consider larger systems. Note that another way of
circumventing the convergence problem in method E might be to add
some potential well to the HF problem for region $A$ to enforce a
stronger localization; although we did not pursue this idea in
this work, we may consider it in the future.

\begin{figure}
\begin{center}\includegraphics[%
  clip,
  height=7cm,
  keepaspectratio]{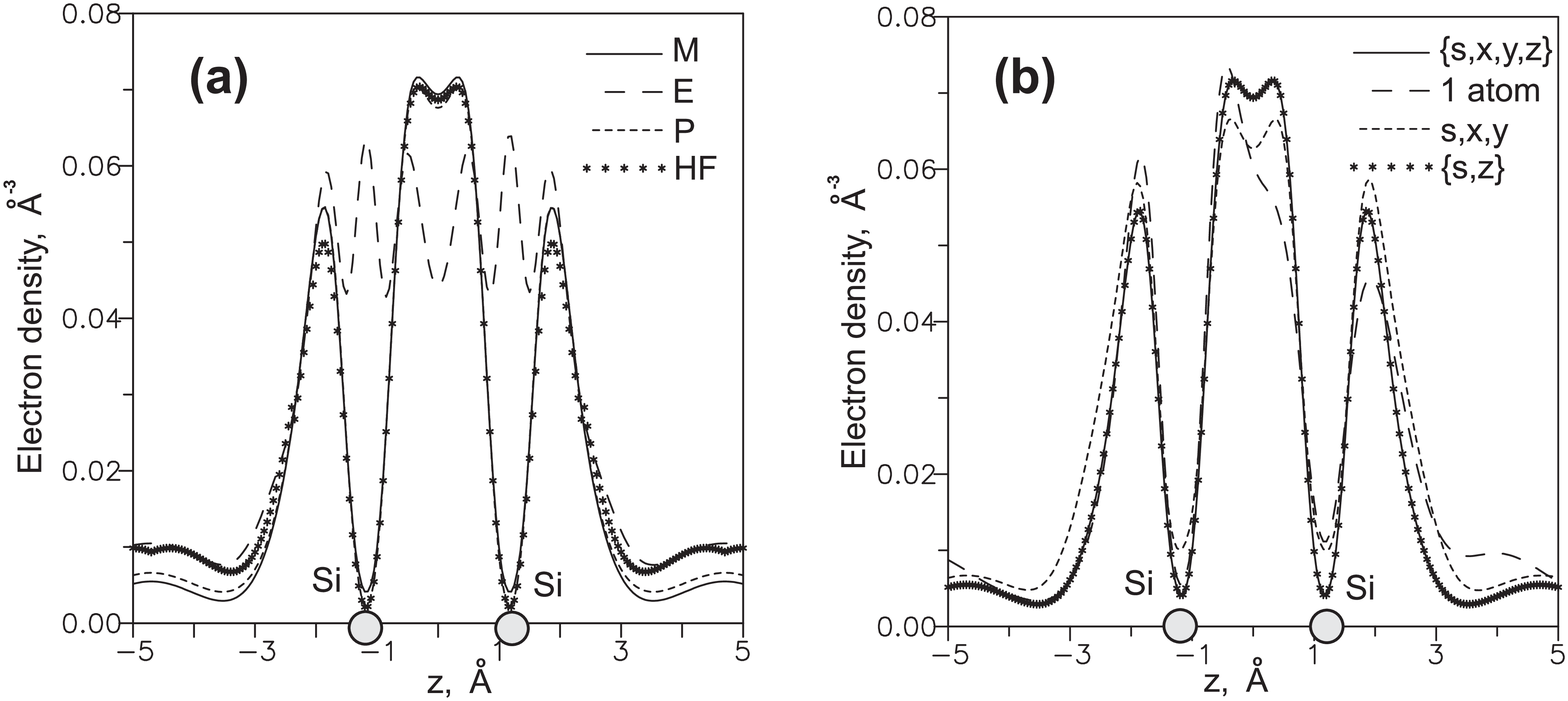}\end{center}

\caption{Electron densities of the Si crystal constructed from
LMOs obtained using different methods based on the largest Si
cluster. All densities are shown along the central Si-Si bond.
Positions of the two Si atoms are also indicated. (a) $\rho_{sz}$
based on $s$ and $p_{z}$ types of AOs using methods M (solid
line), P (small dashes) and E (long dashes) are compared with the
HF density calculated from the middle of the largest cluster
(stars); (b) $\widetilde{\rho}_{1}$ - all AOs centered on a
\emph{single} central Si atom (long dashes);
$\widetilde{\rho}_{sz}$ (stars), $\widetilde{\rho}_{sxy}$ (small
dashes) and $\widetilde{\rho}_{sxyz}$ (solid line) were obtained
using $\{ s,p_{z}\}$, $\{ s,p_{x},p_{y}\}$ and $\{
s,p_{x},p_{y},p_{z}\}$ types of AOs centered on \emph{both}
central Si atoms, respectively; method M was used for all cases in
(b). \label{Fig:comparison-Si}}
\end{figure}

To construct the electron density $\widetilde{\rho}(\mathbf{r})$
of the whole Si crystal we need other three LMOs assigned to the
same primitive cell. These are obtained by appropriate rotations
and displacements of the central bond LMO considered above. The
LMOs corresponding to other crystal cells are then obtained by
applying appropriate lattice translations. The electron densities
$\widetilde{\rho}(\mathbf{r})$ obtained using the three
localization methods for the largest cluster have been thoroughly
compared with the reference density, $\rho(\mathbf{r})$, obtained
by translating the central part of the same cluster. It has been
found that comparison of the densities along the central Si-Si
bond, as shown in Fig. \ref{Fig:comparison-Si} (a), reflects well
the extent in which the densities match with each other.  %
 One can see that both methods M and P lead to the electron density
of the Si crystal which is very close to the reference density
$\rho(\mathbf{r})$. However, the density
$\widetilde{\rho}(\mathbf{r})$ calculated from the LMOs obtained
using method E shows an unphysical oscillatory behavior which is
due to their poor localization and thus an insufficient cluster
size used to construct them.

\section{Discussion and conclusions}

We have seen in the previous section that for an ionic system such
as MgO all localization procedures give identical results and do
not require large cluster sizes. This is because a natural
localization takes place in those systems. In fact we find that
practically identical LMOs can be obtained for MgO using methods M
and P if in addition to the AOs centered on the central O atom one
also adds AOs of any of the nearest Mg atoms to define region $A$.
This means that these localization procedures are sufficiently
flexible in terms of the AOs used to define the localization
regions.

We find that the localization procedures for covalent system with
strong hybridization in its chemical bonding are more sensitive to
the choice of the localization region and the particular
localization method. We have seen above that method E fails for
this system since the LMOs it produces are not sufficiently
localized within the cluster sizes we use.

\begin{figure}
\begin{center}\includegraphics[%
  clip,
  height=6cm,
  keepaspectratio]{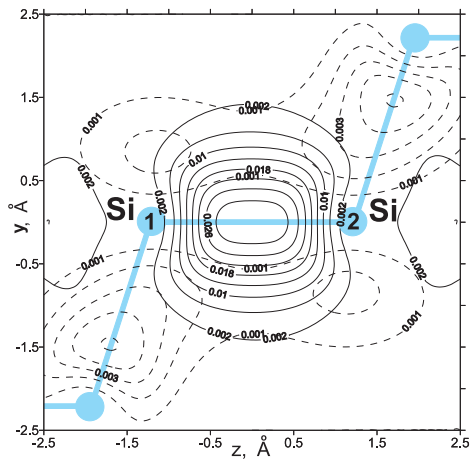}\end{center}

\caption{Contour plots of the LMO partial densities
$\widetilde{\varphi}^{2}(\mathbf{r)}$ obtained using method M for
the two choices of region $A$: (i) all AOs of the two central Si
atoms (solid line) and (ii) only $\{ s,p_{x},p_{y}\}$ AOs (dashed
lines). The plots are calculated in the plane passing through the
Si atoms and two of their nearest neighbors (all four atoms are
indicated). \label{Fig:Si-stupid-orbital}}
\end{figure}

We have also experimented with localization methods M and P by
trying to use different definitions of region $A$ in order to find
the LMO. First of all, we used all AOs on the two central Si
atoms. Similarly to the case of MgO, in this case an LMO
practically identical to the one which we calculated using only
$s$ and $p_{z}$ types AOs was obtained. The $\widetilde{\rho}$
densities calculated in these two cases are also the same as is
demonstrated in Fig. \ref{Fig:comparison-Si} (b) by the good
matching between $\widetilde{\rho}\equiv\widetilde{\rho}_{sxyz}$
and $\widetilde{\rho}\equiv\widetilde{\rho}_{sz}$. The two methods
succeeded since the chemical bonding is essentially correctly
reproduced by either choice of region $A$. This is confirmed by
the contour plot of the partial density
$\widetilde{\varphi}^{2}(\mathbf{r)}$ associated with the LMO and
calculated using all AOs centered on the two central Si atoms: as
shown in Fig. \ref{Fig:Si-stupid-orbital},
the LMO essentially corresponds to the Si-Si bond.%

In contrast, when assuming a wrong character of the chemical
bonding in Si, we obtained LMOs which were either not very well
localized or had completely unexpected (unphysical) spatial
distribution. For instance, assuming atomic character of the
chemical bonding, we attempted to use AOs on a \emph{single} Si
atom to define the single localization region in the primitive
cell. This assumption gave four LMOs similarly to the MgO case. We
find, however, that these LMOs become much less localized and, as
a result, the constructed electron density
$\widetilde{\rho}\equiv\widetilde{\rho}_{1}$ is very different
from $\widetilde{\rho}_{sz}$ as is obvious from Fig.
\ref{Fig:comparison-Si} (b). Another example of an {}``unwise''
choice of the localization region is to use only $s$, $p_{x}$ and
$p_{y}$ AOs of the two central Si atoms to define region $A$
(recall that the two Si atoms are positioned along the $z$ axis).
In this case the total electron density
$\widetilde{\rho}\equiv\widetilde{\rho}_{sxy}$ also shown in Fig.
\ref{Fig:comparison-Si} (b) somewhat differs from
$\widetilde{\rho}_{sz}$, but, at the same time, reproduces all its
main features. However, the LMO itself appears to have a
completely different spatial distribution (see Fig.
\ref{Fig:Si-stupid-orbital}): it is not anymore localized on the
Si-Si bond, but instead was found to be delocalized over a large
volume around it. This explains why the total density
$\widetilde{\rho}_{sxy}$ was found quite different from the
reference one: much larger cluster should be considered to
accommodate fully the LMO obtained using this particular choice of
region $A$. Therefore, the proper choice of the localization
regions which reflect the chemistry of the given crystal results
in more localized orbitals and thus much smaller cluster sizes
needed to construct them.

Concluding, a simple method based on a cluster approach was
suggested in order to construct localized molecular orbitals
(LMOs) for a periodic solid. Our method does not require usage of
periodic codes, is thus much easier to implement in practice and
also, in addition, can also be applied to nonperiodic systems. The
work in this direction is presently in progress and will be
published elsewhere.

Several localization procedures were analyzed and two crystals
were considered in detail (MgO and Si) which corresponds to the
two cases of extreme types of chemical bonding - ionic and
covalent. We find that two localization procedures considered, one
based on the Mulliken populations (method M) and another on a
projection operator (method P), give well localized orbitals with
the expected conventional meaning adopted in chemistry using
already quite moderate cluster sizes. The third procedure, based
on the minimization of the HF energy of a structural element (one
or two atoms) demonstrated a much slower convergence with the
cluster size and is found to be also computationally expensive.

Two cases considered here, MgO and Si crystals, have a well known
type of chemical bonding and thus the choice of the localization
regions in these two cases was obvious. At the same time, we find
that there is a certain degree of flexibility in choosing the
localization regions and this can be exploited in the cases of
more complicated (e.g. intermediate) types of chemical bonding.
This work is being done in our laboratory at present and will be a
matter for future publications.

\section*{Acknowledgments}

We are extremely grateful to I. V. Abarenkov and A. Shluger for a
number of useful and stimulating discussions during his stay in
London. O.D. would also like to acknowledge the financial support
from the Leverhulme Trust (grant F/07134/S) which made this work
possible.

\end{document}